\documentclass[sigconf]{acmart}
\renewcommand\footnotetextcopyrightpermission[1]{}
\acmConference{}
\acmYear{}

\usepackage[font=small,labelfont=bf]{caption}
\usepackage{hyperref}
\usepackage{tikz}
\usepackage{amsmath}
\usepackage{filecontents}
\usepackage{footnote}
\usepackage{diagbox}
\usepackage{listings}
\usepackage{booktabs}
\usepackage{multirow}
\usepackage[ruled,vlined]{algorithm2e}
\usepackage{subfigure}
\usepackage{tabularx}
\usepackage{url}
\usepackage{amsfonts}
\usepackage{algorithmic}
\usepackage{graphicx}
\usepackage{textcomp}
\usepackage{xcolor}
\usepackage[skins]{tcolorbox}
\usepackage{verbatim}
 
\usepackage{enumitem}
\usepackage{pifont}
\usepackage{listings}
\usepackage[framemethod=tikz]{mdframed}
\usepackage{balance}
\usepackage{lipsum} 
\usepackage{indentfirst}

\def\etalnocm{\emph{et al.}\xspace}
\def\etal{\emph{et al.,}\xspace}

\def\BibTeX{{\rm B\kern-.05em{\sc i\kern-.025em b}\kern-.08em
    T\kern-.1667em\lower.7ex\hbox{E}\kern-.125emX}}
\begin{document}

\title{Dissecting Malware in the Wild} 

\author{Hamish Spencer}
\affiliation{%
  \institution{University of Adelaide}
  \country{Australia}
}

\author{Wei Wang}
\affiliation{%
  \institution{University of Adelaide}
  \country{Australia}
}
\author{Ruoxi Sun}
\affiliation{%
  \institution{University of Adelaide}
  \country{Australia}
}

\author{Minhui Xue}
\affiliation{%
  \institution{University of Adelaide}
  \country{Australia}
}

\renewcommand{\shortauthors}{H. Spencer, W. Wang, R. Sun, and M. Xue}
 \renewcommand \authors{Hamish Spencer, Wei Wang, Ruoxi Sun, and Minhui Xue}

\begin{abstract}
With the increasingly rapid development of new malicious computer software by bad faith actors, both commercial and research-oriented antivirus detectors have come to make greater use of machine learning tactics to identify such malware as harmful before end users are exposed to their effects. This, in turn, has spurred the development of tools that allow for known malware to be manipulated such that they can evade being classified as dangerous by these machine learning-based detectors, while retaining their malicious functionality. These manipulations function by applying a set of changes that can be made to Windows programs that result in a different file structure and signature without altering the software’s capabilities. Various proposals have been made for the most effective way of applying these alterations to input malware to deceive static malware detectors; the purpose of this research is to examine these proposals and test their implementations to determine which tactics tend to generate the most successful attacks. \par
\end{abstract}

\maketitle

\section{Introduction}
With the increase of mobile applications and Internet of Things technology in daily life, users are beginning to worry about the security and privacy of software~\cite{sun2021empirical,Feng2021SnipuzzBF,sun2020quality,sun2020venuetrace,sun2021understanding}. 
Despite improving antivirus technology, malware-related exploits remain one of the biggest security problems in computing today, with over 7.2 billion attacks being reported in 2019 alone~\cite{song2020mab}. The issue of malicious Windows executables is of particular concern, as thousands of new samples of harmful programs continue to be uploaded to the VirusTotal online file analyzer each day \cite{demetrio2021functionality}. To address the growing number and complexity of such programs, malware classifiers in both academia and the antivirus industry have quickly adopted machine learning as one of the most popular methods of identifying malicious software \cite{chen2018automated}. \par
This machine learning oriented approach offers several key benefits over traditional methods like a purely signature-based comparison over a database of known samples. For one, bad faith actors do not typically have knowledge of the training techniques or parameters of the machine learning model used in commercial antivirus software, forcing them to perform black-box attacks and so making it more difficult for them to exploit the algorithm with the aim of causing a misclassification \cite{hu2017generating}. Furthermore, machine learning models function by extracting features such as instruction sequences and section names from input data and comparing them to the same features found in known malware samples \cite{kolosnjaji2018adversarial}. This fact makes such models more effective in discriminating between harmful and benign content when it comes to samples that have not been previously seen by the classifier. \par
However, research has also demonstrated that these machine learning models can themselves be exploited by a category of attacks known as adversarial attacks~\cite{shahpasand2019adversarial,chen2019can,wen2021great,li2021hidden,xu2021explainability,li2020invisible,chan2021breaking}. The basis of such attacks is that even the most miniscule of modifications to the file structure of a malware sample, if carefully selected, can bring about misclassification by the machine learning algorithm, even though the program retains its malicious functionality. A malicious program that can be modified to deceive a targeted classifier in this way is known as an adversarial example. \par
Much research has been conducted into methods of most effectively crafting these adversarial examples to deceive well-known malware classifiers in both white box and black box scenarios~\cite{wang2021exposing}. Algorithms for making adversarial attacks often incorporate machine learning-based tactics themselves by experimenting with modifying different semantic aspects of the samples they are trained on and gradually learning which changes tend to be the most effective \cite{grosse2017adversarial}. Implementations of this method have seen some preliminary success in evading harmful classifications by malware detectors, particularly open-source models where white box attacks are viable. However, there is still much room for further research and optimization. \par
The motivation behind this paper is to compare the techniques used in these different methods for generating adversarial examples. By documenting which machine learning tactics and parameters, file modifications, and sampling algorithms tend to generate the most success in exploiting common antivirus programs, these existing models can be optimized to craft adversarial examples even more consistently. In turn, this outcome serves to expose the flaws in commonly available malware classifiers and allows the authors of these programs to understand and fix their vulnerabilities to adversarial attacks, providing end users with greater security against malicious software in general. \par

\section{Background}
To better understand the existing solutions to the problem of creating adversarial examples for popular malware classifiers, we conducted a literature review of relevant, existing research papers. Several of these papers were particularly of interest because they achieved results that demonstrated clear success in deceiving machine learning models into misclassifying malware as legitimate software. \par
Demetrio and Biggio conducted research into different practical manipulations that could be applied to alter the file structure of Windows binaries to evoke a misclassification from machine learning-based antivirus programs \cite{demetrio2021secml}. Their research led to the development of a Python library, known as \emph{secml-malware}, containing tools for generating adversarial examples for a given set of malware samples. The library contains a variety of both white box and black box attacks that each make small, semantic changes to the input binary like deleting the DOS header, shifting the executable content, and padding the executable with random bytes, with the aim of deceiving malware classifiers that rely upon these features to return an accurate identification \cite{demetrio2021secml}. Uniquely, the module also includes GAMMA attacks that pad unused sections of the binary with known goodware (benign content) to mislead antivirus programs into assuming that the malware is legitimate software and classifying it as such. \par
The attacks were tested against a deep neural network-based malware classifier, \emph{MalConv}. Through use of the pytorch machine learning optimizer, the model was able to significantly reduce detection rate after repeated iterations of attacking the classifier. The most effective attacks were those that deleted, extended, or otherwise modified the DOS header in some form, with the classifier reaching detection levels of less than 30\% after 25 iterations \cite{demetrio2021secml}. The black box attacks were noticeably less effective than white box attacks, since the model lacked information about the best direction to take to exploit the classifier. A noticeable exception were the GAMMA attacks, where the low detection rate of the white box attacks could be reached by injecting data from goodware programs, deceiving the \emph{MalConv} model. \par
Another noteworthy paper was a proposal by Song~\etal~ for a reinforcement learning framework for generating adversarial examples for vulnerable classifiers, implemented in another Python library \emph{MAB-Malware} \cite{song2020mab}. It differentiates itself from similar frameworks like \emph{secml-malware} on several counts. For one, it models adversarial example generation in a stateful fashion, meaning that the order of the various possible modifications to the input malware is not considered in training the model. This is beneficial because most of the actions being made in crafting adversarial examples are miniscule changes such as deleting headers and modifying section titles, which are totally independent from one another. Other frameworks that choose to treat the problem as a stateful one, where the order of operations is significant, will see a large increase in learning difficulty, as the number of possible action combinations becomes much larger. The other major improvement of \emph{MAB-Malware} over other models is that when it successfully makes an adversarial attack, any operations that were not necessary for this success are removed prior to points being assigned by the training algorithm \cite{song2020mab}. This further reduces learning time and increases productivity, as actions that did not contribute to crafting an adversarial example are no longer rewarded by the model. \par
On being tested with a set of input malware against the \emph{MalConv} neural network, \emph{MAB-Malware} was able to achieve a detection rate of less than 3\% within 50 iterations of black box attacks \cite{song2020mab}. Of these evasions, the successful adversarial examples were overwhelmingly generated by appending goodware content to the end of the executable, much like the GAMMA attacks in the \emph{secml-malware} library. The further improved evasion rate over those attacks can likely be attributed to \emph{MAB-Malware}’s improved machine learning model. It is worth noting that this exploit was not as successful against three commercial antivirus programs that were also tested, with the framework failing to reach less than a 50\% detection rate against any of them within a reasonable number of iterations – potential methods for optimizing against these classifiers is an area that could be further explored \cite{song2020mab}. \par

\section{Methodology}
The methodology developed for this project was designed with the intention of investigating into the machine learning models and parameters that are most likely to be consistently effective in generating successful adversarial attacks against typical static malware classifiers. These models will be tested not just once, but incrementally over repeated iterations to ensure they have apt time to learn from a given data set the most effective modifications they can make on input malware to deceive antivirus programs. This extensive training process will also ascertain that the models are tested against a wide variety of different malicious executables and that the results are not skewed by any outliers in the input, making the final outcome as accurate as possible to the actual evasion rates these models would produce in practical scenarios. \par

\subsection{Frameworks}
The preliminary stage of the project will involve recreating the results discussed in the two papers discovered during the literature review. This means making use of the \emph{secml-malware} and \emph{MAB-Malware} libraries, running them against appropriate input data, and drawing conclusions from the results. To ensure that the results are as applicable as possible to the problem of crafting adversarial attacks in practical scenarios, it is necessary that this testing process is conducted in a black box manner, where the machine learning model has no knowledge of the internal implementation of the classifier that it is attacking \cite{demetrio2021secml}. For the same reason, the number of iterations the frameworks can make of crafting an attack and improving the action model based on the classifier’s response should be limited to a reasonable number, as it is not realistic for these models to be able to make an unlimited number of attacks in a non-testing environment. Song~\etal~ propose 50 as a reasonable limit on the number of iterations for practical adversarial example generation \cite{song2020mab}. \par

\begin{figure}[h]
\centering
\includegraphics[width=0.45\textwidth]{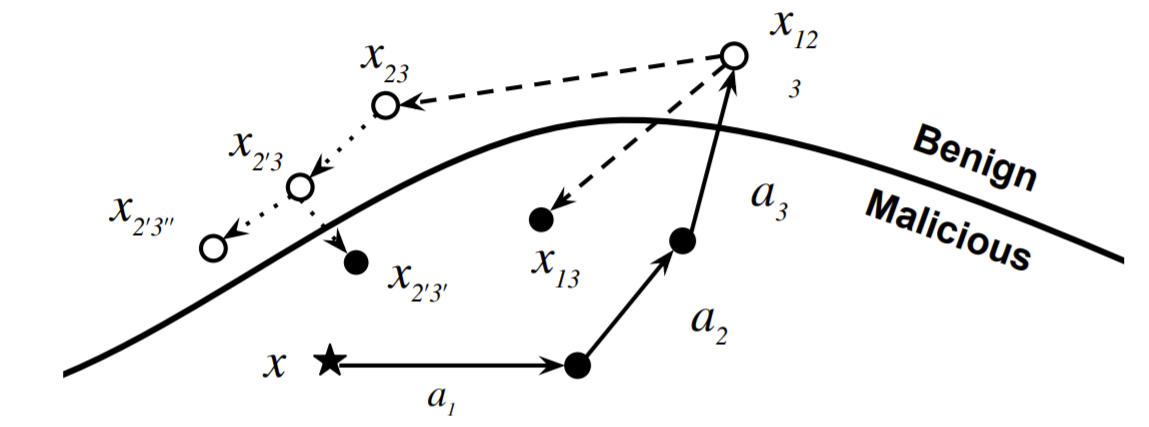}
\caption{\emph{MAB-Malware} action minimization process, adapted from Song~\etalnocm~ \cite{song2020mab}}
\label{minimize}
\end{figure}

The primary way in which the effectiveness of different machine learning features and parameters in generating adversarial examples will be assessed is in comparing the differences in this regard between the two Python libraries being tested. For example, the \emph{secml-malware} framework contains only a simple set of modifications that can be made to the structure of input malware executables \cite{demetrio2021secml}. Whenever an adversarial example is generated, all of the modifications that were involved in changing the executable are assigned points, making them more likely to be used in future adversarial attacks. On the other hand, the \emph{MAB-Malware} framework makes use of a more complicated algorithm for this purpose, by treating the task as a multi-armed bandit (MAB) problem \cite{song2020mab}. Namely, it makes use of the action minimization algorithm shown in figure \ref{minimize}, where inessential actions to the formulation of an adversarial attack are identified and not assigned reward points for that iteration. By examining the results of the two frameworks, we will be able to determine which of these approaches tends to generate more successful exploits. \par

The evasion rates of different machine learning parameters, sampling algorithms, and use of state by the two frameworks will also be considered on an individual basis to assess the importance of these factors in reaching a lower detection rate within fewer iterations by reducing the training time of the model. It will also be important to confirm that the frameworks still consistently retain the original behavior of input malware, as malicious binaries are useless if their harmful behavior is not retained. \par

\subsection{Modifications}
One of the primary factors that determines a good adversarial example generation algorithm is the types of modifications it makes to input executables \cite{hu2017generating}. Each of the two frameworks we are testing have a distinct set of these actions they can take on a given sample in an attempt to transform it into an evasive one. In table \ref{actions} we have compiled a list of such actions that preliminary research suggests have been the most effective in formulating adversarial examples on smaller sample sets, along with which libraries implement them. \par

Each of these modifications makes a different, semantic change that has no effect on the malicious functionality of a program, but can cause an antivirus program to misclassify it as harmless content. The \emph{Edit DOS} and \emph{Extend DOS} actions both manipulate the contents of the typically unused DOS header, which only remains in Windows binaries for retro compatibility \cite{demetrio2021secml}. The former does so by simply injecting random content in its place (retaining only a few key fields that are required to execute the program), while the latter shifts the real DOS header by a specified amount of bytes and injects content before it, taking advantage of the ability to modify an offset within the DOS header to effectively shift the contents of the main program itself. \par

\begin{table}[h]
\centering
\begin{tabular}{ | c | c | c | }
\hline
 \textbf{Action} & \textbf{\emph{secml-malware}} & \textbf{\emph{MAB-Malware}} \\ 
 \hline
 Edit DOS & Yes & No \\
 Extend DOS & Yes & No \\
 Section Append & No & Yes \\  
 Section Add & Yes & Yes \\  
 Section Rename & No & Yes \\
 Padding & Yes & Yes \\
 Code Randomization & No & Yes \\
 \hline
\end{tabular}
\caption{Executable modifications implemented by each library}
\label{actions}
\end{table}

The \emph{Section Append} and \emph{Section Add} attacks modify the input binary by adding random bytes to the unused space of an existing section or a new section entirely, respectively \cite{song2020mab}. \emph{Section Rename} makes a similar modification to the name of the executable sections, which do not have any practical effect on the behaviour of the program, by changing them to titles associated with known goodware code. The \emph{Padding} attack functions by appending bytes directly at the end of the executable file, which are sourced from benign programs much like in the \emph{Section Rename} action, although the \emph{secml-malware} also offers the option to just pad with completely random bytes \cite{demetrio2021secml}. The \emph{Code Randomization} manipulation is unique in that it randomly rearranges the machine language instructions within the binary in such a way that the semantic behaviour of the executable remains unchanged. The implementation of this attack require particular caution to ensure that the malicious effect of the program is retained. \par

The aim of this research is to examine each of these action types both on an individual basis and in combination with other actions that can be taken against input binaries. In doing so we could then make conclusions about which kinds of modifications tend to be the most efficient in forcing a misclassification from common malware classifiers against malicious software. This would allow for the synthesis of an algorithm for generating adversarial examples that is more optimal than existing solutions. \par

In the process of planning this project, it is important that we also consider any contingencies that could arise in undertaking the research methodology. One particular issue that is predicted as being a possible concern is the process of identifying which of the properties of each of the frameworks is contributing to their higher or lower successful evasion rates against different malware samples. Based on how the results are presented within the research papers, we assume that the problem of decoupling the classifier detection rate of factors such as attack types, sampling algorithms, and white box or black box learning from the detection rate of the frameworks as a whole is a trivial one. \par

However, if it does transpire that there is no clear method within the controls of one or both of the \emph{MAB-Malware} and \emph{secml-malware} frameworks for assessing the effectiveness of these properties, on an individual basis, in generating adversarial examples, then it would be much more difficult for us to make meaningful recommendations for improving the evasion rate of the models, as there is no clear method of discerning which factors are responsible for a particular frameworks success or lack of success. \par

To address this issue, it will likely be necessary for us to edit the source code of the adversarial example generation programs, and temporarily restrict their set of features to the particular ones that we wish to test. For example, in the \emph{secml-malware} framework, if we want to determine how successful the Padding manipulation is in making an adversarial attack, we would need to manually delete all of the other attack types from the model, train it as usual, and then compare the results to those generated by other attack types in the same fashion. \par

If this version of the model tend to experience more success in evading malware classifiers than others, then we could conclude that the Padding attack is an effective one. While this process would be tedious, it could potentially be necessary for us to make adequate conclusions about the types of models that more consistently generate adversarial examples. Having this contingency plan in place is crucial for us to ensure the research is able to continue even if the analysis tools provided with the two frameworks being tested turns out to be lacking. \par

\subsection{Input Data}
A dataset containing around 2,500 samples of known malicious Windows binaries has been obtained, with the intention of feeding it as input to the training algorithms of both frameworks to ensure they both undertake their learning in the same fashion and have sufficient information to optimize the process of making minor modifications to given binaries to produce an adversarial example. This sample set consists of executables that were identified by typical, commercial antivirus programs employed by everyday users. As such, this input data will be useful in allowing us to generate results that will be as relevant as possible to the practical field of malware classification. 

In considering the potential contingencies that could occur in conducting this research, another one of the biggest issues that is predicted to possibly arise is this Windows malware dataset turning out to be in some way unsuitable for the goals of this project. This could occur, for example, if some of the samples are not initially being recognized as malicious by the malware classifiers we are using, if they do not make use of the legacy DOS headers that are exploited in several of the adversarial attacks, or if the sample size turns out to be too small for meaningful training of the models to take place. We plan to address this issue by attempting to locate a more appropriate malware dataset that does contain the features required to analyze the effectiveness of the two adversarial example generation frameworks being tested. For this purpose, online tools like the VirusTotal file scanner provide access to various datasets of user-uploaded content that have been identified as malicious by one or more antivirus product. These malware sets are therefore employed as potential replacements for the existing data we plan on using to train the \emph{MAB-Malware} and \emph{secml-malware} models, if they prove to be in some form inadequate. \par

\subsection{Data Preparation}
Before beginning the process of running the model on the input data, it is important that the dataset is processed and prepared so as to be suitable for training the machine learning models being tested. In the case of our main dataset, we will need to ascertain that the input binaries are in fact being classified as malware by the \emph{MalConv} classifier to begin with, as evoking a misclassification is not a meaningful result if an antivirus engine could not already determine that a program is malicious.

\subsection{Malware Classifiers}
It is important to ensure that the conditions in which the two frameworks are being tested are identical so that their strengths and weaknesses in crafting adversarial examples are assessed on a fair basis. To achieve this, both models will be tested against a pre-trained copy of the \emph{MalConv} neural network, ensuring the quality of malware detection is the same in each case. \emph{MalConv} was the ideal choice for this research because it is an advanced malware classifier that is also open-source and well reputed in the research community \cite{raff2017malware}. This makes it easy for us to investigate into where our attacks might be failing and how we could optimize them to a greater extent to evade the classifier.

To further explore the performance of the models in different scenarios we will additionally experiment with the \emph{EMBER} (Elastic Malware Benchmark for Empowering Researchers) malware dataset, which provides a structured set of features collated from over a million malicious executables for the purposes of training adversarial example generation \cite{anderson2018ember}. To even more effectively mirror the issues associated with generating adversarial examples in a real-life situation, the learning models could be also tested against commercial antivirus programs, instead of just research-oriented malware classifiers like \emph{MalConv}. \par

\subsection{Final Model}
Once the testing environment has been reproduced and the new outcomes are recorded, the next stage will involve analyzing these results and identifying, on this basis, the most effective strategies for composing adversarial attacks. Of particular importance is considering the different executable modifications like editing sections, appending additional content, and randomizing the instruction sequence, and determining which of these changes are most significant in generating successful adversarial examples. \par

The final stage of the project will involve taking these observations about the significance of different features of the machine learning-based adversarial example generation models and using them to make recommendations for the building of a more efficient framework that can achieve a higher evasion rate over a smaller number of iterations against common malware classifiers. To achieve this, the most effective executable modifications will be combined with the fastest training model and sampling algorithm, as well as other proposed features such as stateless adversarial example generation, culling of redundant actions, and use of goodware content in byte injection, if these features have been demonstrated in testing to lead to more productive attacks. \par

Once these ideal attributes have been detailed, we could then fork the source code of the \emph{secml-malware} and \emph{MAB-Malware} frameworks and edit them to add features that bring them closer to this optimal model. In turn, these forked frameworks could be tested further with the expectation of higher evasion rates, with the aim of confirming our hypothesis about the most efficient features in adversarial example generation. \par

\section{Experimental Setup}
In order to accurately evaluate the different models being considered throughout this research, it is essential that the input data being fed to the frameworks is equivalent and has been properly filtered for unsuitable or missing entries. A realistic number of iterations for which the model should run and optimize the adversarial examples over needs to be determined and kept consistent among frameworks. Additionally, the metrics for which the effectiveness of the different machine learning models, action types, and sampling methods can be compared and evaluated so as to produce some credible conclusions also needs to be outlined. \par

\subsection{Data Processing}
The sample set of around 2,500 input malware samples discussed previously proved to be infeasible for testing as a whole within the scope of this research. The primary issue with the dataset was that there were too many executables for it to be processed within a reasonable amount of time. This was especially pertinent in the case of attacks adapted from the \emph{MAB-Malware} library, likely owing to the framework's complicated action minimization algorithm discussed previously. For each adversarial example generated by the framework, the action minimizer must iterate over every major action performed on the original executable and attempt to remove it \cite{song2020mab}. If after removing the action's effects from the evasive sample, the file still evokes a misclassification from the malware classifier, then that action is considered redundant and can be removed without being assigned reward points. If removing the action makes the sample no longer evasive, then it can be considered essential to the generation of the adversarial attack, and given the points as usual. However, the algorithm must still iterate over all of the minor actions within that larger, essential action and attempt to remove them in a similar fashion to determine which parts of the action are truly essential \cite{song2020mab}. \par

These minor actions can include modifications as small as adding or removing one byte at the end of a section within the executable. Since all combinations of these tiny changes need to be generated and passed to the \emph{MalConv} classifier, the attacks implemented in \emph{MAB-Malware} quickly become very costly to process. Although the major actions can be run individually on a dataset, this process of optimizing out unnecessary minor actions still occurs, which is the main bottleneck on performance. Even with access to high-performance computing resources, the largest number of samples that could be processed using the \emph{MAB-Malware} library for periods of sustained testing was 100. Since we wanted to test all of the attacks and modification types in the same environment and with the same input data, the testing for the \emph{secml-malware} actions also had to be limited to input sizes of just 100. \par

\subsection{Confidence Levels}

\begin{figure}[h]
\centering
\includegraphics[width=0.45\textwidth]{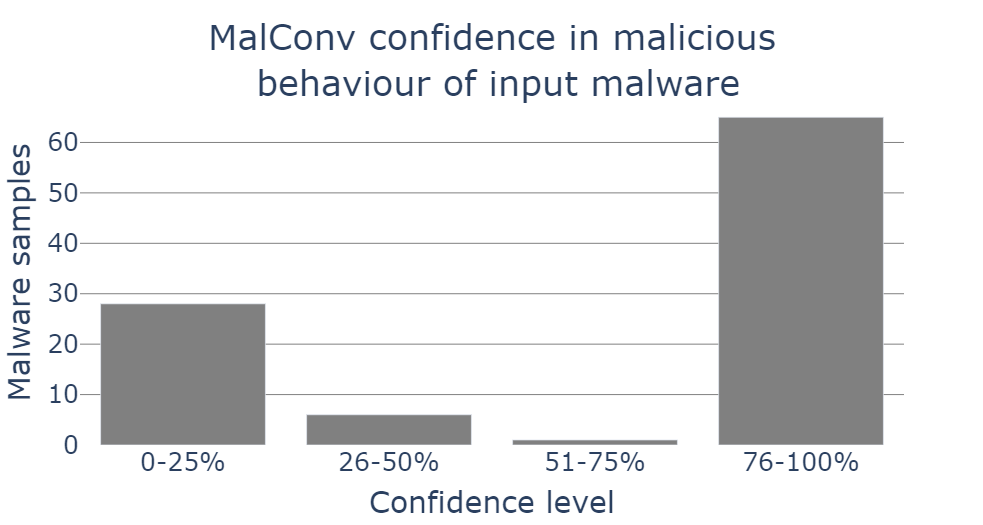}
\caption{\emph{MalConv} confidence levels in malicious behaviour over a dataset of 100 Windows malware samples}
\label{conf}
\end{figure}

Another essential consideration that needed to be made in conducting this research was ensuring that the input malware samples being fed to the attack scripts are actually being recognized initially as malicious. If they are not,  then the samples being considered evasive after actions are applied is an uninteresting outcome, as the classifier already could not determine that it was processing malware. Unlike some antivirus engines that return a simple true or false result as to whether the input is determined to be malicious, the \emph{MalConv} classifier that we are using outputs a confidence rating on how likely the file is to be malware. The confidence levels returned for each executable passed to the classifier from our dataset of 100 executables are shown in figure \ref{conf}. \par

\begin{figure}[h]
\centering
\includegraphics[width=0.45\textwidth]{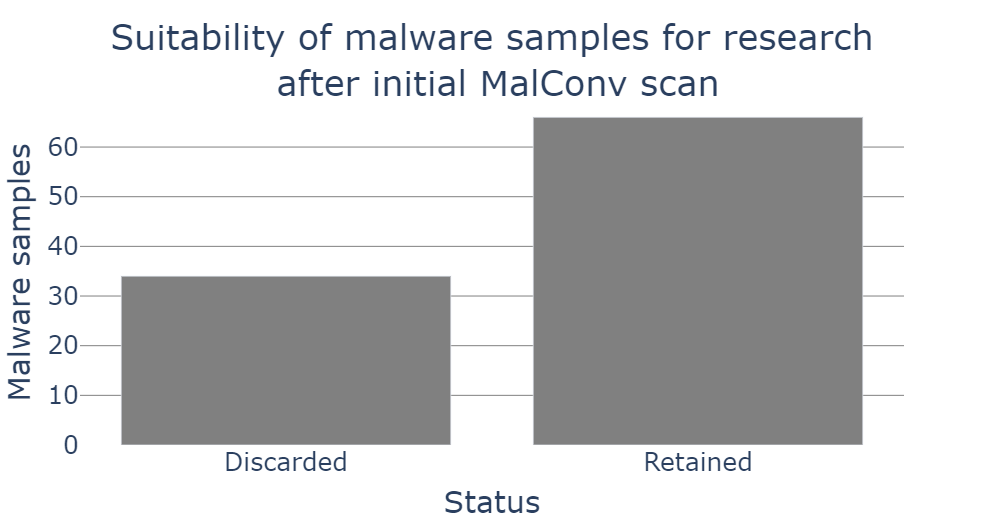}
\caption{Sample executables removed from the malicious dataset due to low confidence levels}
\label{suit}
\end{figure}

To adapt to this system, it is necessary for us to determine a confidence level at which we will consider a sample initially undetected and exclude it from the data set. It would also be useful to determine such a number because it simplifies the process of determining at what point an adversarial attack has made a sample into an evasive one. We ultimately settled on 50\% as the baseline of confidence for malicious input, as this rating suggests that the classifier expects that the executable is more likely to be malware than not. If a sample with a confidence level any lower than this is chosen, we will not be able to gather as much useful information from its evasion rate over repeated iterations, so we chose to remove these from the input. As illustrated in figure \ref{suit}, this resulted in a small portion of our dataset being removed as unsuitable for research. \par

\subsection{Evaluation of Algorithm Performance}
To make conclusions regarding the effectiveness of the different adversarial example generation techniques, it was necessary for us to define some metrics by which we could determine what constituted better attack performance. We determined that the ideal number of attack iterations over which it was reasonable to assess the effectiveness of an algorithm in creating an evasive sample for a given executable was 20. While this number could realistically be increased, from examining the results of previous research~\cite{demetrio2021functionality}, 20 iterations tends to be the point at which an adversarial machine learning-based model begins to hit diminishing returns with further optimizing the action sequence used to evade a classifier \cite{demetrio2021secml}. Furthermore, if the number of iterations were any less, the framework may not have sufficient time to learn from its mistakes and generate an adversarial example, making the evasion rate lower than it would be in a practical scenario. \par

To compare the algorithmic performance of the different action types being tested throughout this research, it will be necessary to also formulate a method of isolating them from other factors and determining their standalone evasion rate. This is a straightforward process with the attacks implemented in the \emph{secml-malware} library, as each action type is provided as a separate module. As such, it is a trivial task to test the modification types individually by importing them and passing them the same input data, and then comparing the subsequent evasion rates. This is also achievable with the  \emph{MAB-Malware} framework, as it supports a configuration file that allows for certain features and actions to be enabled and disabled as needed, although this does not allow for the same kind of fine-tuned control available with importing each attack individually. \par

\section{Results}
All of the relevant modification types listed in table \ref{actions} were tested by taking the implementations of the corresponding attacks from the \emph{secml-malware} and \emph{MAB-Malware} libraries and running them over 20 iterations for each executable in the malware input. The final evasion rate achieved for each sample was recorded using a Python script, along with some general statistics about the entire input like the average number of iterations required for an adversarial example to be generated. This output was then converted into Python dataframes for plotting with the \emph{matplotlib} library, as well as piped to a comma-separated values file for further analysis with Microsoft Excel. Different machine learning parameters like the sampling method were also altered and tested for various values to provide some insight into the optimization of these settings. \par

\subsection{Action Types}
\begin{figure}[h]
\centering
\includegraphics[width=0.45\textwidth]{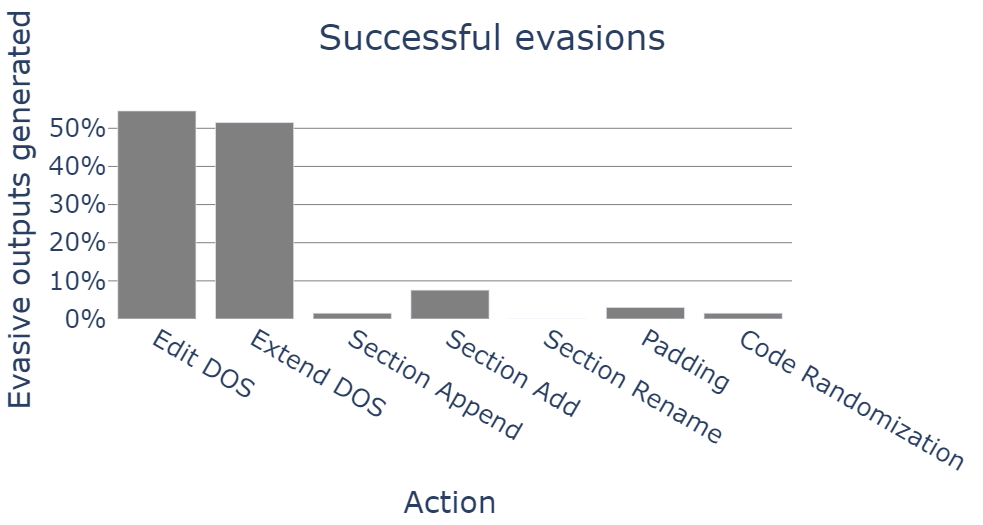}
\caption{Percent of malware samples for each action for which an adversarial example was generated}
\label{success}
\end{figure}

Figure \ref{success} illustrates the relative success rates of each of the different action types in generating an adversarial example for a given piece of sample malware in the input. As can be seen in the graph, the most successful types of attacks tend to be those that manipulate the content of the legacy DOS header. This is likely due to the fact that one of the only aspects of the DOS header that cannot be removed without breaking the executable format is a pointer to the real Windows header \cite{raff2017malware}. By editing the DOS header to modify this pointer's location or destination, the offsets of all of the different sections in the file can effectively be shifted by an arbitrary amount. The fact that this entirely rearranges the layout of the executable, and that the \emph{MalConv} classifier we are using is heavily reliant on structural analysis of files to learn what is and is not malware, offers a possible explanation as to why these DOS attacks cause such a drop in the detection rate \par

\begin{figure}[h]
\centering
\includegraphics[width=0.45\textwidth]{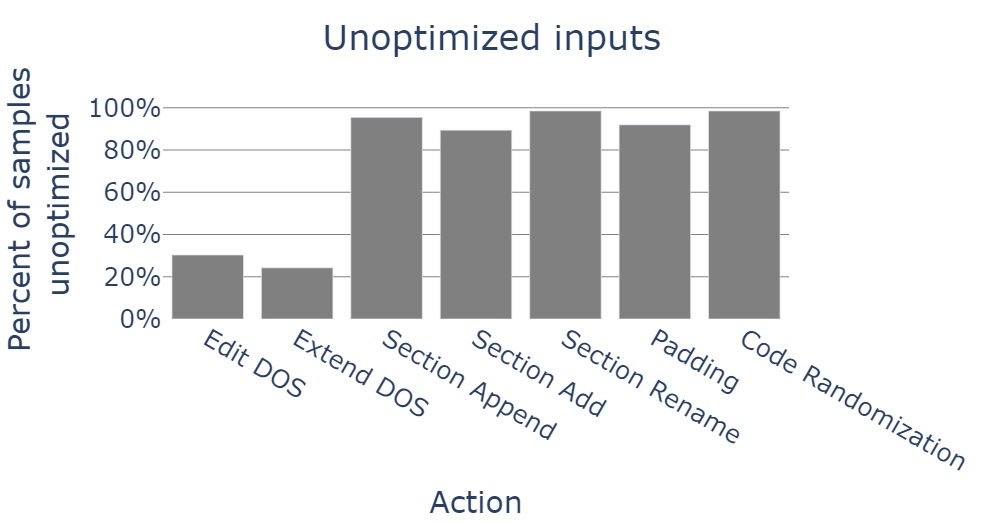}
\caption{Percent of malware samples for each action type for which the MalConv confidence level could not be decreased at all}
\label{unoptimized}
\end{figure}

The graph in figure \ref{unoptimized} shows the portion of samples fed to each attack type that the model could not optimize the \emph{MalConv} malware confidence rate for at all, even after completing 20 iterations of training. Almost all of the samples run with the Section Append and Code Randomization actions fell under this category, demonstrating that the names of the sections within the file and the order of the instruction sequence are not particularly relevant to assessing whether or not a given program is malicious. \par

Interestingly, the Extend DOS attack was the least likely to not be able to make any optimizations to an input malware file, failing to do so in only around 25\% of instances. However, this action type was flawed in other respects. Along with the Section Add attack, it would break the structure of the malicious file in around one in five cases, rendering the executable unusable. From investigating into the issue, however, this can likely be attributed to problems with the detection of header sizes in the \emph{secml-malware} implementation of these attacks, as opposed to a fundamental flaw with how they work. \par

\begin{figure}[h]
\centering
\includegraphics[width=0.45\textwidth]{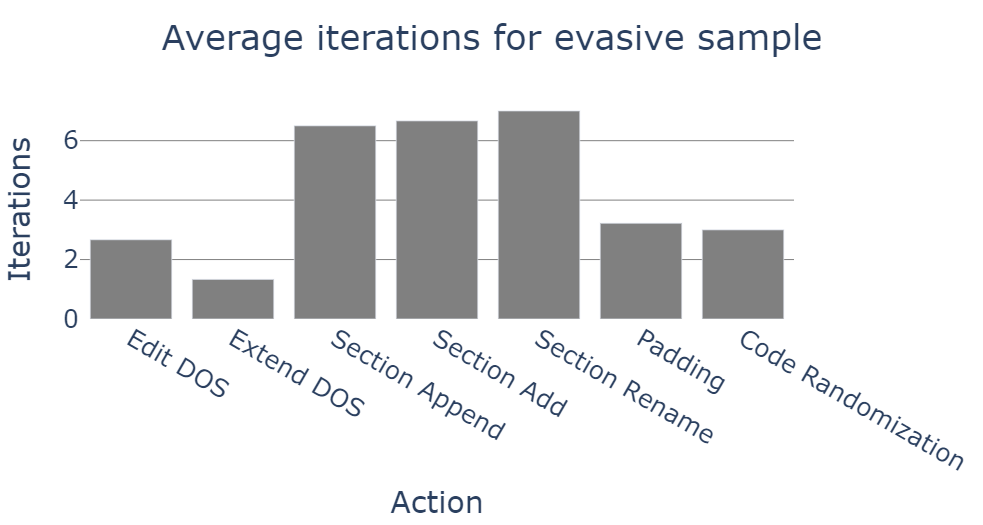}
\caption{Average iterations for each action type required to create an evasive sample, for inputs where an adversarial example was successfully generated}
\label{average}
\end{figure}

Figure \ref{average} illustrates the average iterations required by each attack type to generate an adversarial example, where it did successfully create such an example. Unsurprisingly, the attacks that tended to be more likely to do so in general also managed to achieve this in less iterations of the machine learning model, whereas less efficient action types such as Section Append and Section Rename needed more iterations to learn how to evade the classifier. \par

\begin{figure}[h]
\centering
\caption{Percent of malware samples for each action type for which an adversarial example was created within a single iteration of the model}
\label{immediate}
\end{figure}

Figure \ref{immediate} shows the percent of malware samples for which an evasive sample was generated within a single iteration of the model, without any need for further training. The Extend DOS attack was able to achieve this with around half of all malware samples, likely owing to the fact that it is a simple action that just shifts the main header by a custom number of bytes, so there is not much room for optimization. The Edit DOS attack, despite being more successful overall, was less likely to create an evasive sample on the first iteration, probably because the model needs to learn what kind of bytes are appropriate to insert in the DOS header and where exactly to put them. \par

\subsection{Iterations}
\begin{figure}[h]
\centering
\includegraphics[width=0.45\textwidth]{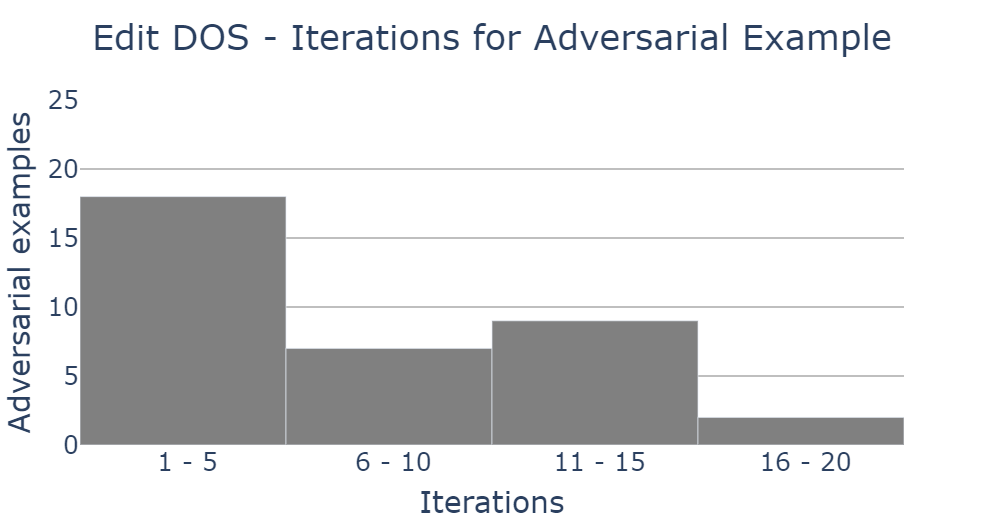}
\caption{Iterations required to create an adversarial example using the Edit DOS attack type, in cases where an evasive sample was successfully generated}
\label{editdos}
\end{figure}

\begin{figure}[h]
\centering
\includegraphics[width=0.45\textwidth]{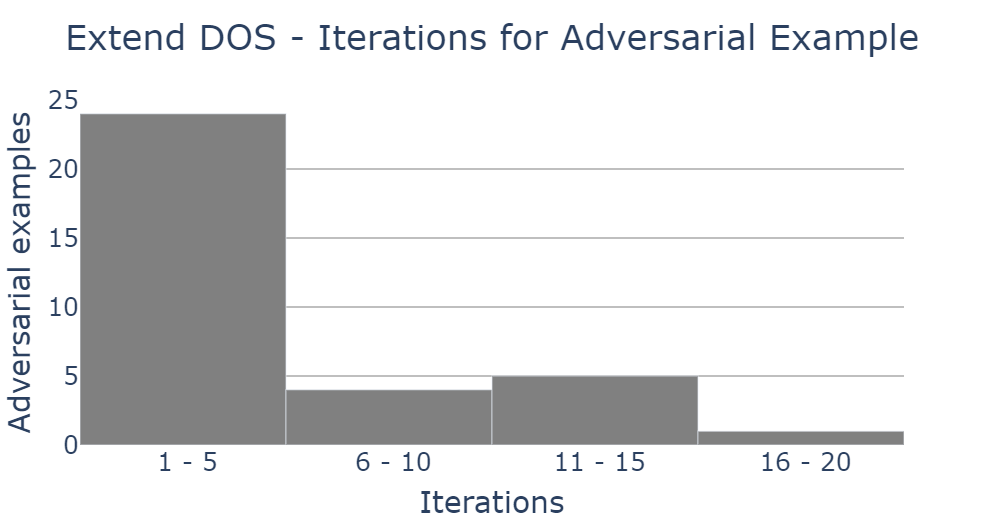}
\caption{Iterations required to create an adversarial example using the Extend DOS attack type, in cases where an evasive sample was successfully generated}
\label{extenddos}
\end{figure}

\begin{figure}[h]
\centering
\includegraphics[width=0.45\textwidth]{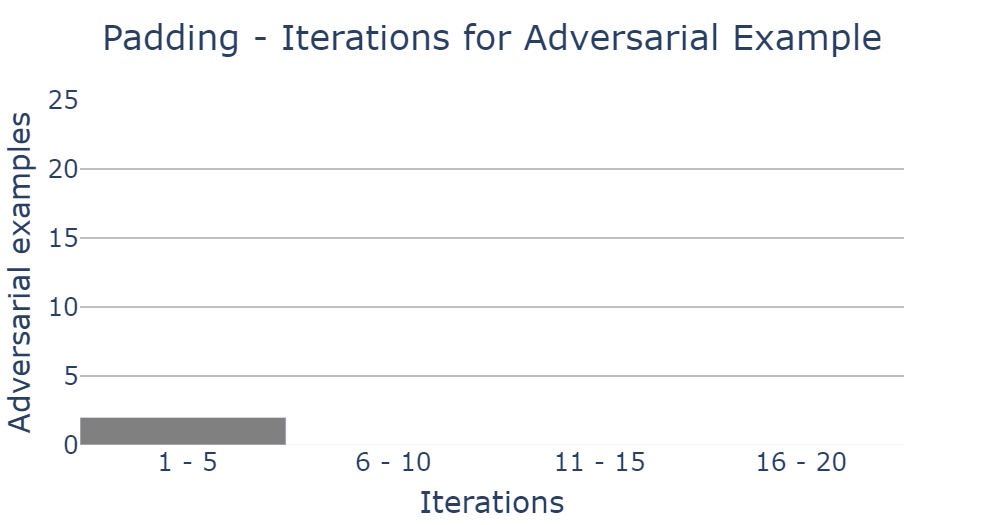}
\caption{Iterations required to create an adversarial example using the Padding attack type, in cases where an evasive sample was successfully generated}
\label{padding}
\end{figure}

\begin{figure}[h]
\centering
\includegraphics[width=0.45\textwidth]{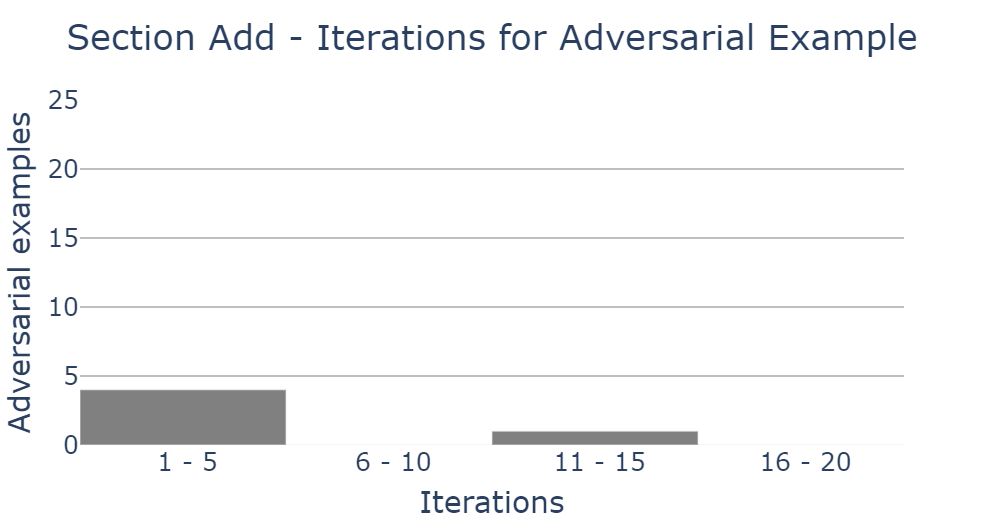}
\caption{Iterations required to create an adversarial example using the Section Add attack type, in cases where an evasive sample was successfully generated}
\label{sectionadd}
\end{figure}

Pictured in figures \ref{editdos}, \ref{extenddos}, \ref{padding}, and \ref{sectionadd} are histograms showing the number of iterations of the machine learning algorithm that was required to create an adversarial example for the Edit DOS, Extend DOS, Padding, and Section Add attacks respectively. The same data could not be gathered for the Section Append, Section Rename, and Code Randomization action types because they are only implemented in the \emph{MAB-Malware} library, which does not offer the same level of detailed information about the status of each individual malware sample being processed as \emph{secml-malware}. \par

Of these attacks, the Edit DOS and Extend DOS actions were the only two that were able to achieve consistent, meaningful success in generating adversarial examples on the given malware dataset. As seen in figure \ref{extenddos}, the Extend DOS attack was able to optimize each input executable to pass a confidence test by the \emph{MalConv} classifier within 1-5 iterations in the vast majority of cases, likely owing to the simplicity of this action type overall. The Edit DOS attack had more variance in the number of iterations required to create an evasive sample, but still only required up to five iterations in around half of all cases. What is most interesting about figures \ref{editdos} and \ref{extenddos} is that both attacks very rarely required any more than 15 iterations to generate an adversarial example, suggesting that the maximum number of iterations could be reduced without having a significant impact on the framework's success rate. \par

Data is sparse for the Padding and Section Add attacks in figures \ref{padding} and \ref{sectionadd}, since these action types achieved very little success in generating adversarial examples in general. However, it is worth noting that the same conclusions hold for these attacks. The vast majority of evasive samples were created within the first five iterations, and none of them needed more than 15 iterations to create. \par

\subsection{Sampling Method}
\begin{figure}[h]
\centering
\includegraphics[width=0.45\textwidth]{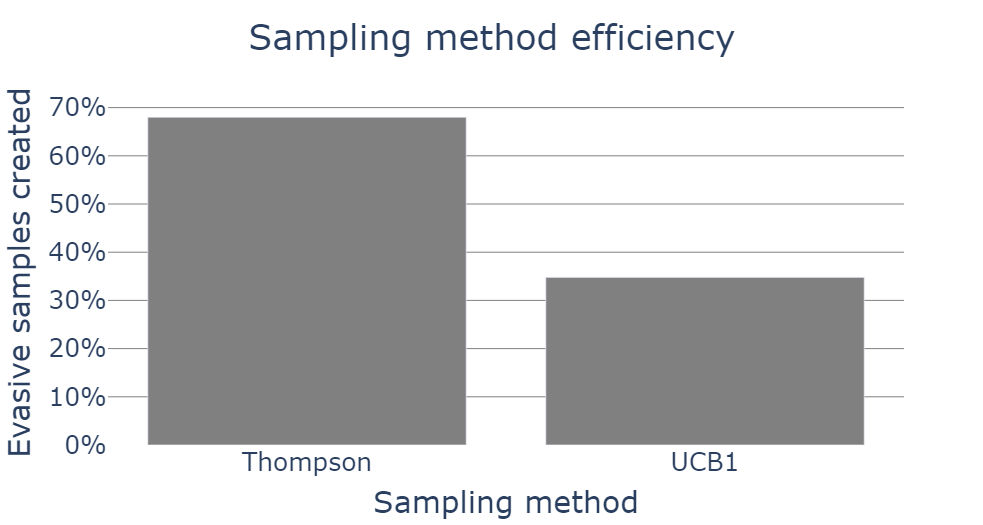}
\caption{Percent of input malware for which an adversarial example could be created for different sampling methods}
\label{sampling}
\end{figure}
As discussed previously, the \emph{MAB-Malware} framework makes use of a complicated machine learning model where evasive samples are created by applying combinations of actions to input malware, and the modifications that contributed to adversarial attacks are assigned points \cite{song2020mab}. These reward points are then considered and used in deciding which action combinations should be applied to future malware samples to create an adversarial example. \par

In testing the \emph{MAB-Malware} machine learning model, two options are available for the sampling method used to select these action combinations to run on the sample set. UCB1 is a simple sampling algorithm that just takes actions with the highest reward point values to create the action combination that should be applied to the current piece of malware being processed \cite{song2020mab}. Thompson sampling, on the other hand, strikes more of a balance between exploration of all of the actions available and exploitation of those known to have been effective on previous samples.

This is achieved through keeping an uncertainty score that begins high and shrinks as more samples are processed by the model. When the uncertainty is high, the framework is more likely to select actions with lower point values, but as it decreases only combinations that have been more successful in the past will tend to be chosen. \par

These two sampling methods were tested for a period of 4,500 iterations over the given malware sample set. As illustrated in figure \ref{sampling}, Thompson sampling was almost twice as effective in creating evasive samples as the more simple UCB1 algorithm. This demonstrates that some level of exploration is necessary in sampling to allow all of the action types a proper opportunity to generate adversarial examples. \par


\section{Conclusion}
The primary conclusions drawn from this research related to the types of actions applied to input malware files that were most effective in creating adversarial examples. We showed that when it came to the \emph{MalConv} malware classifier in particular, evasive samples were most commonly created using attack types that edit the legacy DOS header retained in Windows binaries for retro compatibility. This can be attributed to the presence of a pointer in the DOS header to the rest of the file, which can be manipulated by these attacks to effectively rearrange the entire file structure, a modification that \emph{MalConv} has difficulty dealing with. Actions that only manipulated the section names and content of the executable, as well as the instruction sequence of the assembly code, tended to be less effective in generating evasive samples. \par

The maximum number of iterations allowed for the modifications applied by a particular action on a given sample to be optimized could be reduced to as low as 15, as testing showed that attacks generally experienced diminishing returns beyond this point. The importance of adopting sampling methods that adequately explore all action types available in attempting to create an evasive sample, instead of just picking those that have been the most successful in the past, was also demonstrated. The code of our experiments has been published~\footnote{https://github.com/hspen4/adversarial-attacks}.

Future research in this area could investigate into the possibility of attempting to create evasive samples for commercial antivirus engines, not just \emph{MalConv}. The effectiveness of the \emph{MAB-Malware} action minimizer in optimizing the rewards provided to different actions could also be explored.\par

\bibliographystyle{unsrt}
\bibliography{references}

\end{document}